\documentclass[aps, amsmath, amssymb, aps, pre, twocolumn,groupedaddress,superscriptaddress,nofootinbib,longbibliography]{revtex4-1}
\usepackage{lmodern}

\usepackage{multirow}
\usepackage{verbatim}
\usepackage{graphicx}
\usepackage[config, labelfont={sf,bf}, textfont=sf]{caption,subfig}
\usepackage[suffix=, outdir=./figs/]{epstopdf}

\usepackage{enumitem}
\setlist{noitemsep,topsep=0pt,parsep=0pt,partopsep=0pt}

\usepackage{array}
\usepackage{hhline}
\usepackage{xfrac}
\usepackage[colorlinks,linkcolor=black,citecolor=blue,urlcolor=black]{hyperref}

\newcommand{\xsub}[1]{%
  \mbox{\scriptsize\begin{tabular}{@{}c@{}}#1\end{tabular}}%
}

\DeclareMathOperator{\LEs}{LE}
\DeclareMathOperator{\LCEs}{LCE}

\newtheorem{definition}{Definition}
\newtheorem{theorem}{Theorem}

\usepackage{listings}
\usepackage[usenames,dvipsnames]{color}
\definecolor{MyDarkGreen}{rgb}{0.0,0.4,0.0}
\begin{document}

\title{Finite-time and exact Lyapunov dimension of the Henon map.
}

\author{N.V. Kuznetsov}
\email[]{Corresponding author: nkuznetsov239@gmail.com}
\affiliation{Faculty of Mathematics and Mechanics, St. Petersburg State University,
Peterhof, St. Petersburg, Russia}
\affiliation{Department of Mathematical Information Technology,
University of Jyv\"{a}skyl\"{a}, Jyv\"{a}skyl\"{a}, Finland}
\author{G.A. Leonov}
\affiliation{Faculty of Mathematics and Mechanics, St. Petersburg State University,
Peterhof, St. Petersburg, Russia}
\affiliation{Institute of Problems of Mechanical Engineering RAS, Russia}
\author{T.N. Mokaev}
\affiliation{Faculty of Mathematics and Mechanics, St. Petersburg State University,
Peterhof, St. Petersburg, Russia}

\date{\today}


\begin{abstract}
This work is devoted to further consideration of the H\'{e}non map with negative values of
the shrinking parameter and the study of transient oscillations, multistability,
and possible existence of hidden attractors.
The computation of the finite-time Lyapunov exponents by different algorithms is discussed.
A new adaptive algorithm for the finite-time Lyapunov dimension computation
in studying the dynamics of dimension is used.
Analytical estimates of the Lyapunov dimension using the localization of attractors are given.
A proof of the conjecture on the Lyapunov dimension of self-excited attractors
and derivation of the {exact Lyapunov dimension} formula are revisited.
\end{abstract}

\maketitle
\section{Introduction}
In 1963, American mathematician and meteorologist Edward Lorenz
published an article in which he investigated an approximate
model of the fluid convection in a two-dimensional layer
and discovered its irregular behavior~\cite{Lorenz-1963}.
The analysis of this behavior was closely related to the
phenomenon of chaos, namely to the sensitive dependence on the initial conditions
and the existence in the phase space of the system
of a chaotic attractor with a complex geometrical structure.
Later, this three-dimensional system was called the Lorenz system,
and so far it has great scientific interest~\cite{Tucker-1999,Stewart-2000,LeonovK-2015-AMC}.

In 1969, French mathematician and astronomer, Michel H\'{e}non, showed
that the essential properties of the Lorenz system
(i.e., folding and shrinking of volumes)
can be preserved by means of a
specially chosen sequence of approximating maps~\cite{Henon-1976}.
The Poincar\'{e} map of the Lorenz system was chosen as the reference,
resulting in the following map
$\varphi(x,y) : \mathbb{R}^2 \to \mathbb{R}^2$, where
\begin{equation}\label{sys:henon}
  \varphi(x,\,y) = \left(1 + y - a x^2, \, b x\right),
\end{equation}
$a > a_0 = -\tfrac{(b-1)^2}{4}$ (folding parameter),
and $b \in (0,1)$ (shrinking parameter)
are parameters of mapping.
One usually considers
an equivalent form of the H\'{e}non map (e.g.~\cite{BoichenkoL-2000,Hunt-1996,GrassbergerKM-1989})
\begin{equation}\label{sys:henon2}
  \varphi(x,\,y) = \left(a+by-x^2, \, x\right),
\end{equation}
obtained from~\eqref{sys:henon} by changing the coordinates
$x := a x$, $y := \tfrac{a}{b} y$.
Along with positive values of parameter $b$, considered in~\cite{Henon-1976},
later on one also studies the H\'{e}non system with negative values
$b \in (-1, 0)$~\cite{Feit-1978,Heagy-1992}.

The H\'{e}non map and its various generalizations have attracted the attention of researchers
with its comparative simplicity and the ability to model its dynamics
without integrating differential equations
(see, e.g. \cite{BenedicksC-1991,BenedicksY-1993,GrassbergerKM-1989,SterlingDM-1999,GonchenkoGS-2010,FalcoliniTL-2013,GaliasT-2013}).
Despite being introduced initially as a theoretical transformation,
it also has several physical interpretations~\cite{BihamW-1989,Heagy-1992}.
Further, we will study H\'{e}non map in the form~\eqref{sys:henon2} with $b \neq 0$, $|b| < 1$.

\section{Transient oscillations, attractors and multistability}
Consider a \emph{dynamical system} with discrete time
$\big(\{\varphi^t\}_{t\geq0},(U \subseteq \mathbb{R}^2,||\cdot||) \big)$
generated by the recurrence equation with map \eqref{sys:henon2}
\begin{equation}\label{sys:map}
   u(t+1) = \varphi(u(t)), \ \ u(0) = u_0 \in U, \ t \in \mathbb{N}_0,
\end{equation}
where $\varphi^t = \underbrace{(\varphi \circ \varphi \circ \cdots \circ \varphi)}_{\xsub{\rm $t$ times}}$,
$\varphi^0 = {\rm id}_{\mathbb{R}^2}$,
and $u(t) = (x(t),y(t))$ with $||u_0|| = \sqrt{x_0^2 + y_0^2}$.

\begin{figure}[h]
 \centering
 \includegraphics[width=0.49\textwidth]{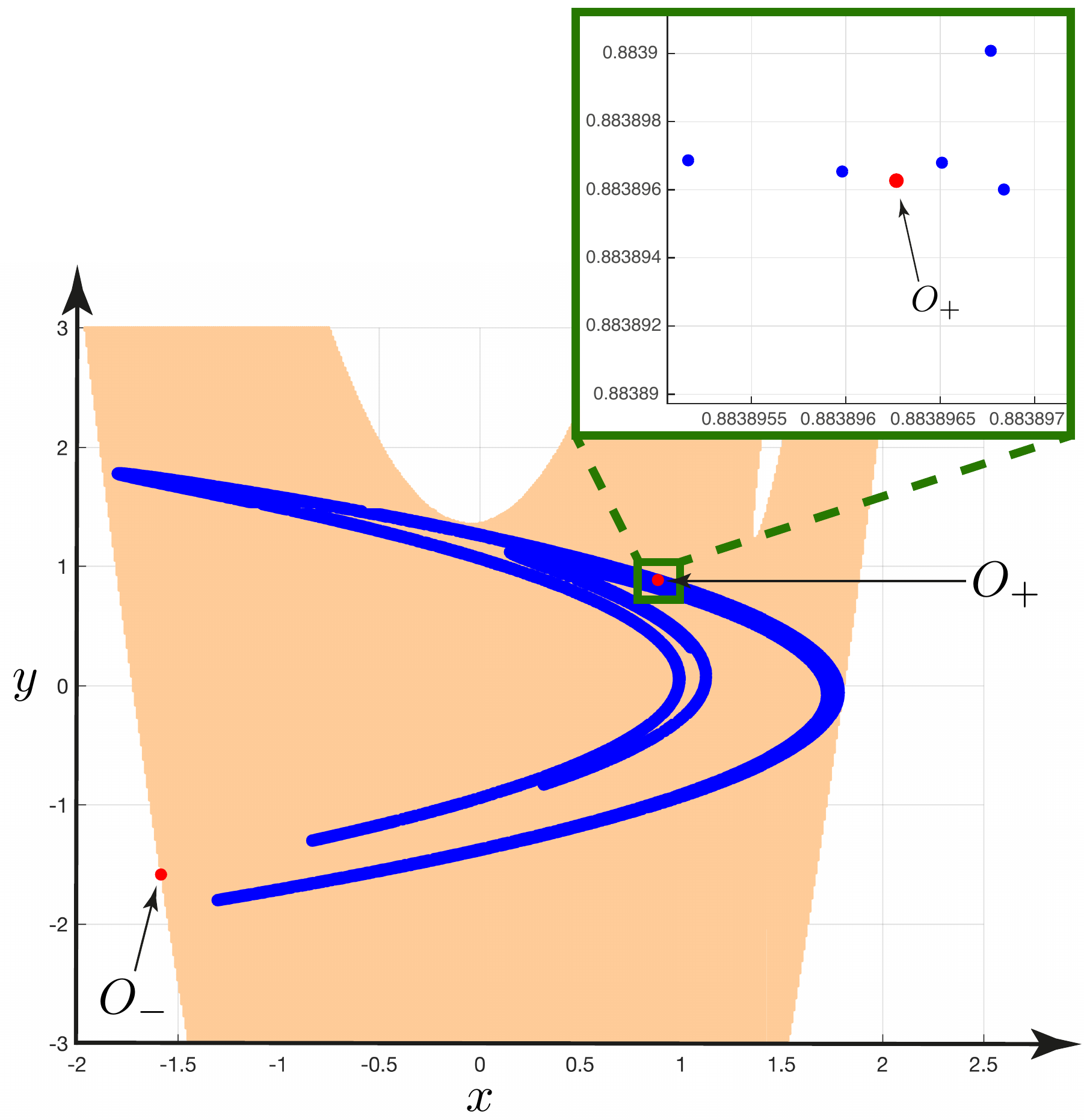}
 \caption{\label{fig:henon:attr:SE}
 Chaotic H\'{e}non attractor (blue) for the parameters $a = 1.4$, $b = 0.3$
 is self-excited with respect to both equilibria $O_\pm$ (red);
 the basin of attraction (orange).
 }
\end{figure}

\begin{figure*}[t]
 \centering
 \subfloat[After the transient process (cyan) the trajectory from
 the $\delta$-vicinity of the equilibrium $O_-$, $\delta = 0.01$
 localizes the chaotic attractor (blue).]{
 \label{fig:henon:attr:NON_HID}
 \includegraphics[width=0.45\textwidth]{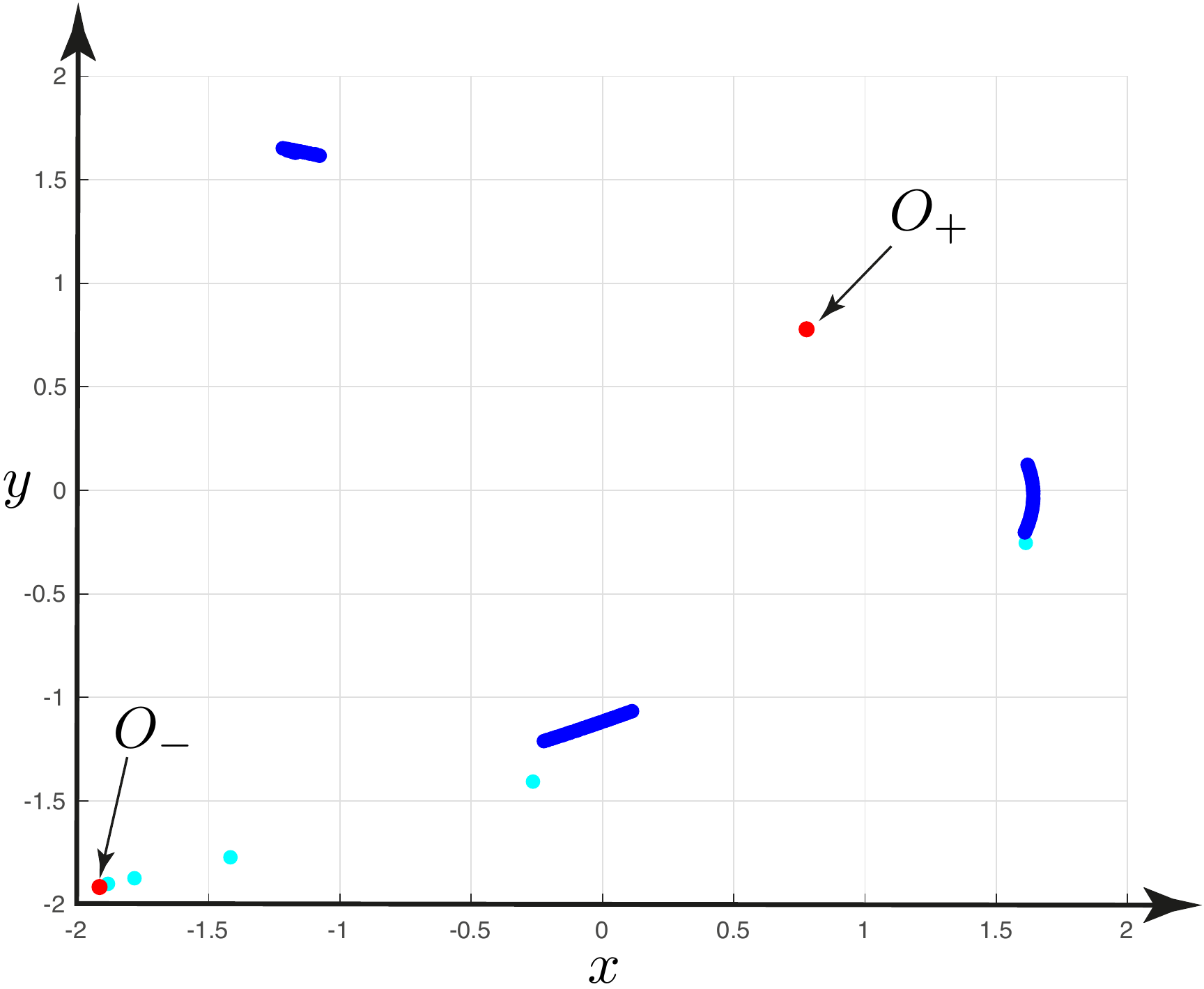}
 }~
 \subfloat[ Self-excited periodic attractor (black) with respect to
 the unstable equilibria $O_\pm$ (red)
 and its basin of attraction (green);
 self-excited chaotic attractor (blue)
 with respect to the unstable equilibrium $O_{-}$ (red),
 its basin of attraction (orange).]{
    \label{fig:henon:attr:HID}
    \includegraphics[width=0.45\textwidth]{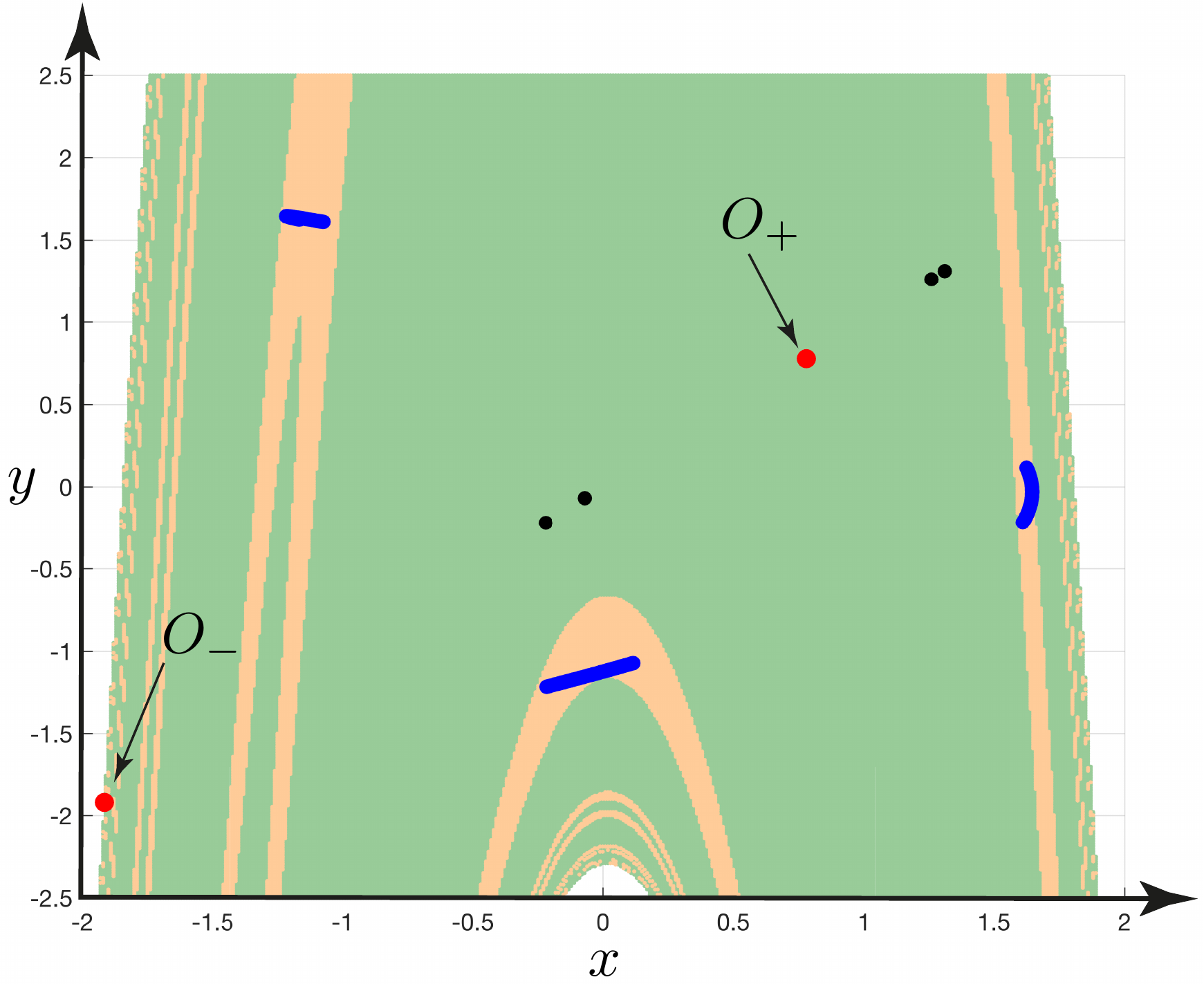}
  }
 \caption{\label{fig:henon:transient}
   Multistability and coexistence of two local attractors
   in the H\'{e}non system
   with $a = 1.49$, $b = -0.138$.
 }
\end{figure*}

The equilibria $O_{\pm} = (x_\pm,x_\pm)$ of the system \eqref{sys:map} exist when $a > a_0$.
Here
\[
  x_\pm=\frac{1}{2} \,\big(b-1\pm\sqrt{(b-1)^2+4a}\big).
\]
The $2 \times 2$ Jacobian matrix is defined as follows:
\begin{equation}\label{eq:jac}
  J(u_0)\!=\!J\big((x_0,y_0)\big)\!=\!D\varphi\big((x_0,y_0)\big) = \left(
          \begin{array}{cc}
            -2x_0 & b \\
            1 & 0 \\
          \end{array}
        \right),
\end{equation}
where $|\det J(u_0)| \equiv |b| <1$, and
has the eigenvalues $\lambda^\pm (u_0) = -x_0 \pm \sqrt{b + x_0^2}$,
and eigenvectors $\nu^\pm (u_0) = 
\left(\begin{array}{c}
  -x_0 \pm \sqrt{b + x_0^2} \\ 1
\end{array}\right)$.
One can check that equilibrium $O_-$ is always unstable
and $O_+$ is unstable when $a > a_1 = \tfrac{3(b-1)^2}{4}$.

In the seminal work~\cite{Henon-1976} for a fixed value $b = 0.3$
it was shown numerically that for $a < a_0$ and $a > a_3 \approx 1.55$
any trajectory $\varphi^t(u_0)$ tends to infinity for all $u_0 \in \mathbb{R}^2$.
For $a_0 < a < a_3$, depending on the initial point $u_0$,
the trajectory $\varphi^t(u_0)$ either tends to infinity, or
reaches the attractor that is a stable equilibrium for $a_0 < a < a_1$,
a periodic orbit for $a_1 < a < a_2 \approx 1.06$, and
a nontrivial chaotic attractor for  $a_2 < a < a_3$.
These numerical experiments imply that there is no global attractor in the
H\'{e}non system for $U = \mathbb{R}^2$.

Computational errors (caused by finite precision arithmetic)
and sensitivity to initial conditions 
allow one to get a reliable visualization of chaotic attractor
by only one pseudo-trajectory computed for a sufficiently large time interval.

\begin{definition}[\cite{KuznetsovLV-2010-IFAC,LeonovKV-2011-PLA,LeonovKV-2012-PhysD,LeonovK-2013-IJBC}]
 An attractor is called a \emph{self-excited attractor}
 if its basin of attraction
 intersects with any open neighborhood of a stationary state (an equilibrium),
 otherwise, it is called a \emph{hidden attractor}.
\end{definition}

For a \emph{self-excited attractor} its basin of attraction
is connected with an unstable equilibrium
and, therefore, self-excited attractors
can be localized numerically by the
\emph{standard computational procedure}
in which after a transient process a trajectory,
started in a neighborhood of an unstable equilibrium (e.g., from a point of its unstable manifold),
is attracted to the state of oscillation and then traces it.
Thus, self-excited attractors can be easily visualized
(e.g. the classical Lorenz and R\"{o}ssler attractors can be visualized
by a trajectory from a vicinity of unstable zero equilibrium).

\begin{figure*}[t]
 \centering
 \subfloat[Transient chaotic set existing for $t = 1,\dots,T \approx 7300$.]{
    \label{fig:henon:transient:attr}
    \includegraphics[width=0.49\textwidth]{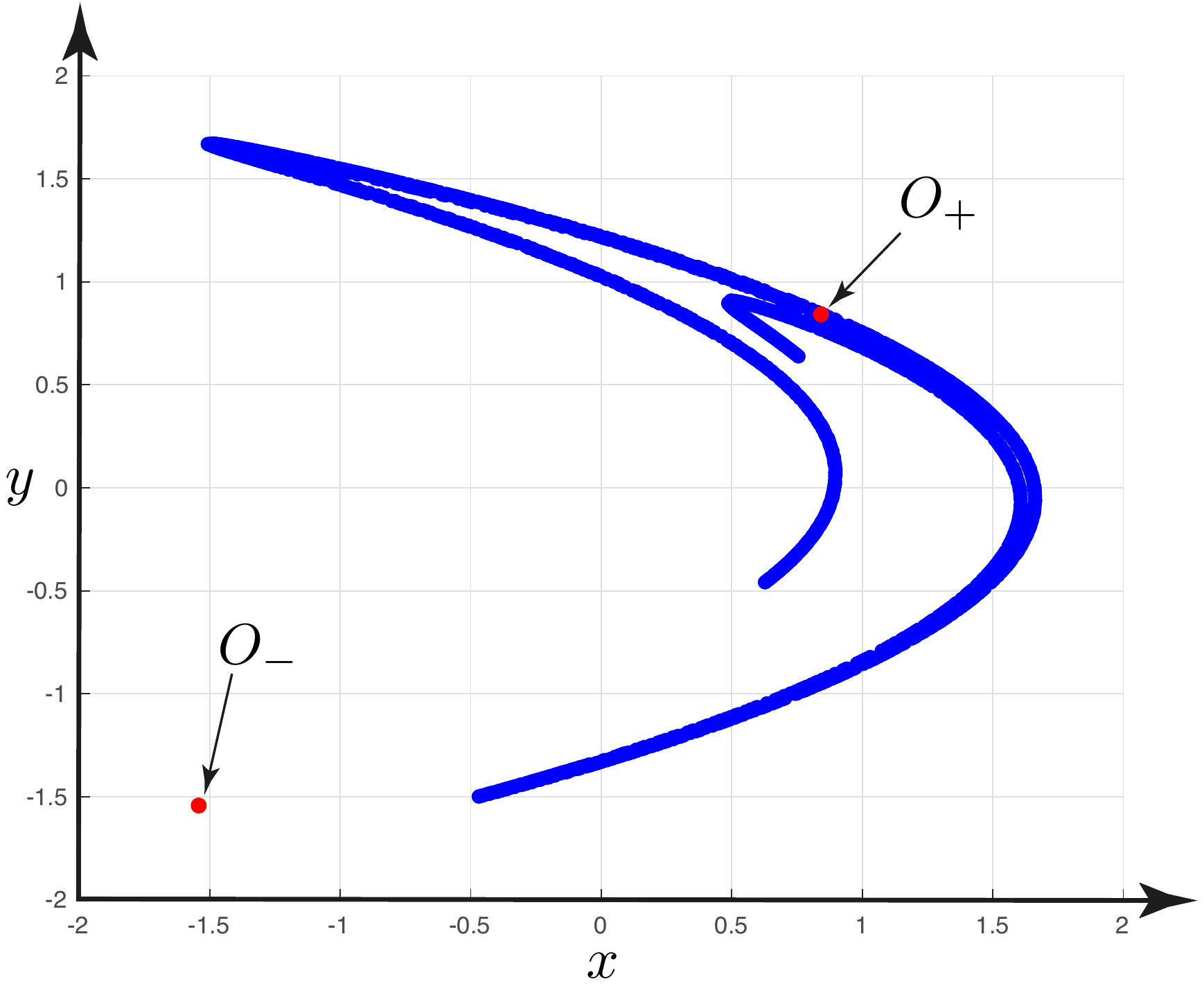}
  }~
  \subfloat[{Period-7 limit cycle, resulting from the collapsed transient set, $t > 7300$.}]{
    \label{fig:henon:transient:LC}
    \includegraphics[width=0.49\textwidth]{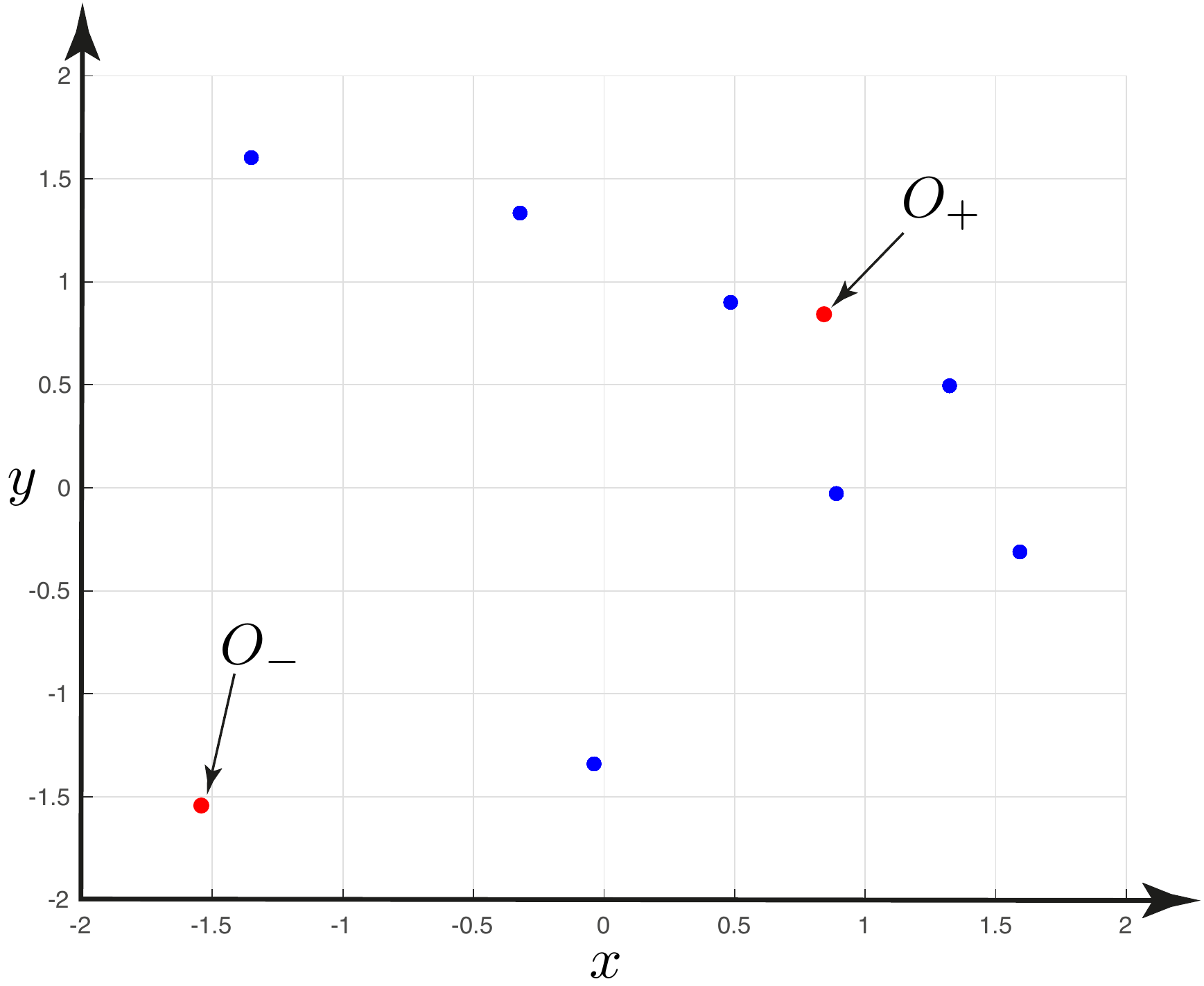}
  }
 \caption{\label{fig:henon:transient}
 For the H\'{e}non system \eqref{sys:map} with parameters $a = 1.29915$, $b = 0.3$
 the evolution of the self-excited transient chaotic set (blue)
 with respect to $O_+$.
 }
\end{figure*}

For $a = 1.4$, $b = 0.3$ the equilibria $O_{\pm}$ are saddles
($|\lambda^\pm(O_\pm)| < 1$, $|\lambda^\mp(O_\pm)| > 1$)
and one can visualize the classical H\'{e}non attractor~\cite{Henon-1976} (see Fig.~\ref{fig:henon:attr:SE})
from the $\delta$-vicinity of $O_{\pm}$ using trajectories with the initial data
\[
  u_0 = O_\pm  \ + \ \delta \frac{\nu^\mp(O_\pm)^*}{||\nu^\mp(O_\pm)||},
  \quad \delta = 0.1.
\]
Similar self-excited attractor can be also obtained
for negative values of parameter $b$
(e.g. for $a = 2.1$, $b = -0.3$ in~\cite{Feit-1978,Heagy-1992}).

In~\cite{DudkowskiPK-2016} the multistability and possible existence of hidden attractor in the H\'{e}non system
$a = 1.49$, $b = -0.138$
were studied by the perpetual point method~\cite{Prasad-2015,DudkowskiJKKLP-2016}.
For these values of parameters $|\lambda^+(O_-)| > 1$, $|\lambda^-(O_-)| < 1$
and from the initial point
\[
  u_0 = O_- + \delta \frac{\nu^+(O_-)^*}{||\nu^+(O_-)||}, \quad \delta \leq 0.01
\]
on the unstable manifold of the saddle $O_-$, defined by the eigenvector $\nu^+(O_-)$,
a chaotic attractor in Fig.~\ref{fig:henon:attr:NON_HID} can be visualized.
Thus, the chaotic attractor obtained in~\cite{DudkowskiPK-2016}
is a self-excited attractor with respect to $O_-$.
In addition, a self-excited periodic attractor can be visualized
from vicinity of $O_+$.
See coexisting self-excited periodic and chaotic attractors and their basins of attraction
in Fig.~\ref{fig:henon:attr:HID}.

\begin{figure*}[t]
 \centering
 \subfloat[Parameters $a = 0.98$, $b = 0.4415$: period-8 (blue), period-12 (cyan), and period-20 (pink)
 orbits.]{
    \label{fig:henon:multistab:3sinks}
    \includegraphics[width=0.33\textwidth]{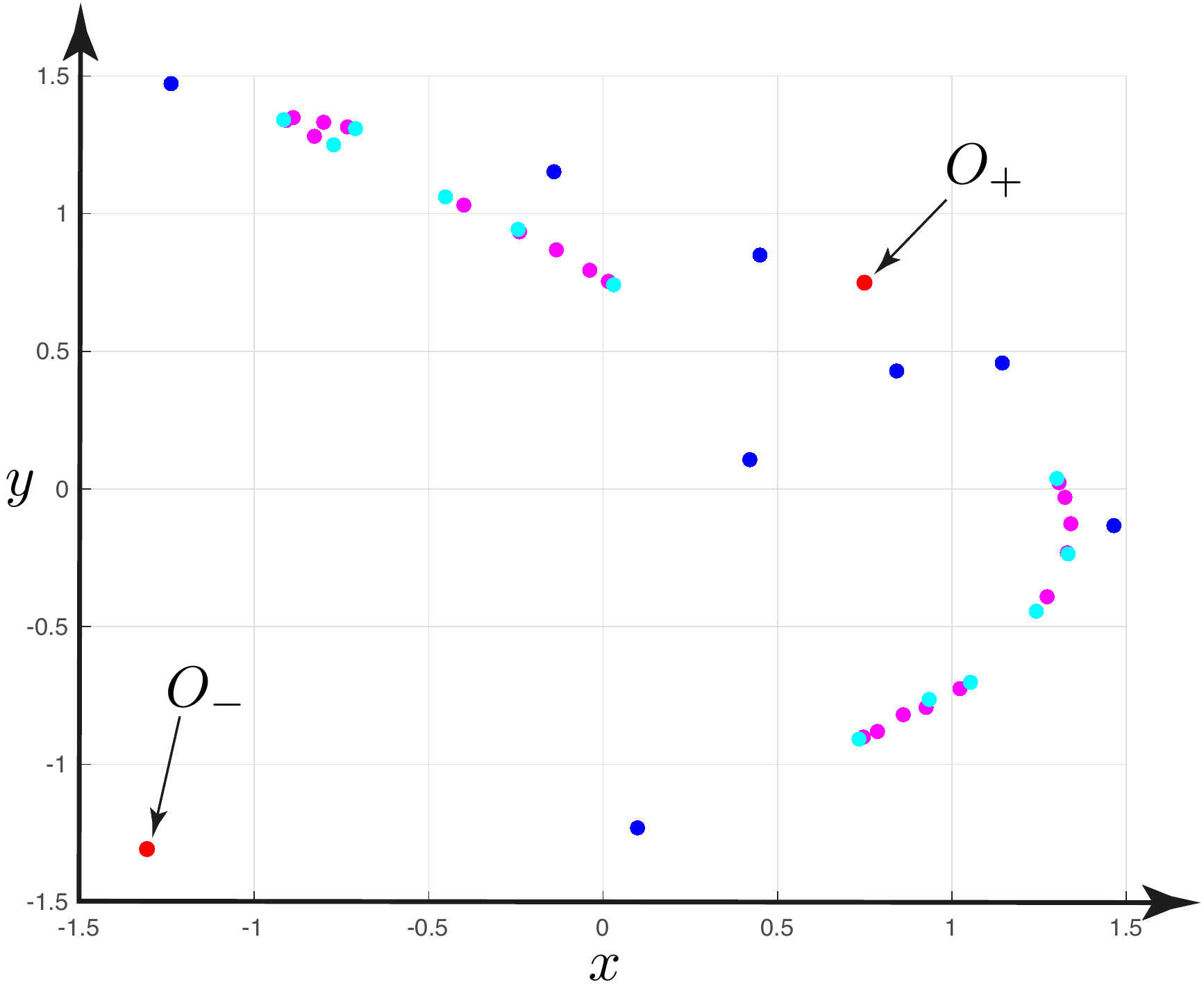}
  }~
  \subfloat[Parameters $a = 0.972$, $b = 0.4575$: period-12 (cyan), period-16 (pink) orbits,
  and chaotic attractor (blue).]{
    \label{fig:henon:multistab:2sinks-1chaos}
    \includegraphics[width=0.33\textwidth]{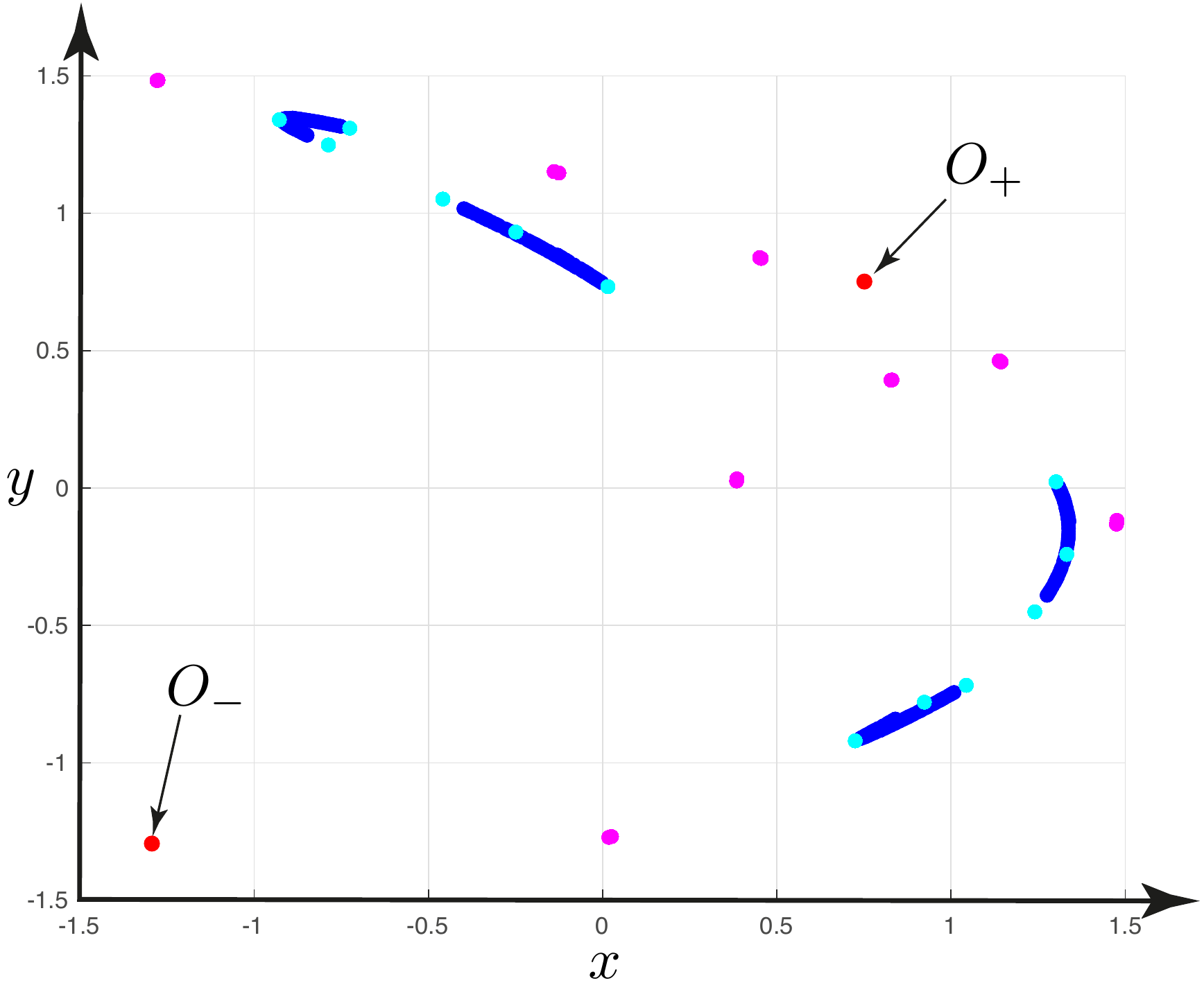}
  }~
  \subfloat[Parameters $a = 0.97$, $b = 0.466$: period-8 orbit (cyan), and two chaotic attractors (blue, pink).]{
    \label{fig:henon:multistab:1sink-2chaos}
    \includegraphics[width=0.33\textwidth]{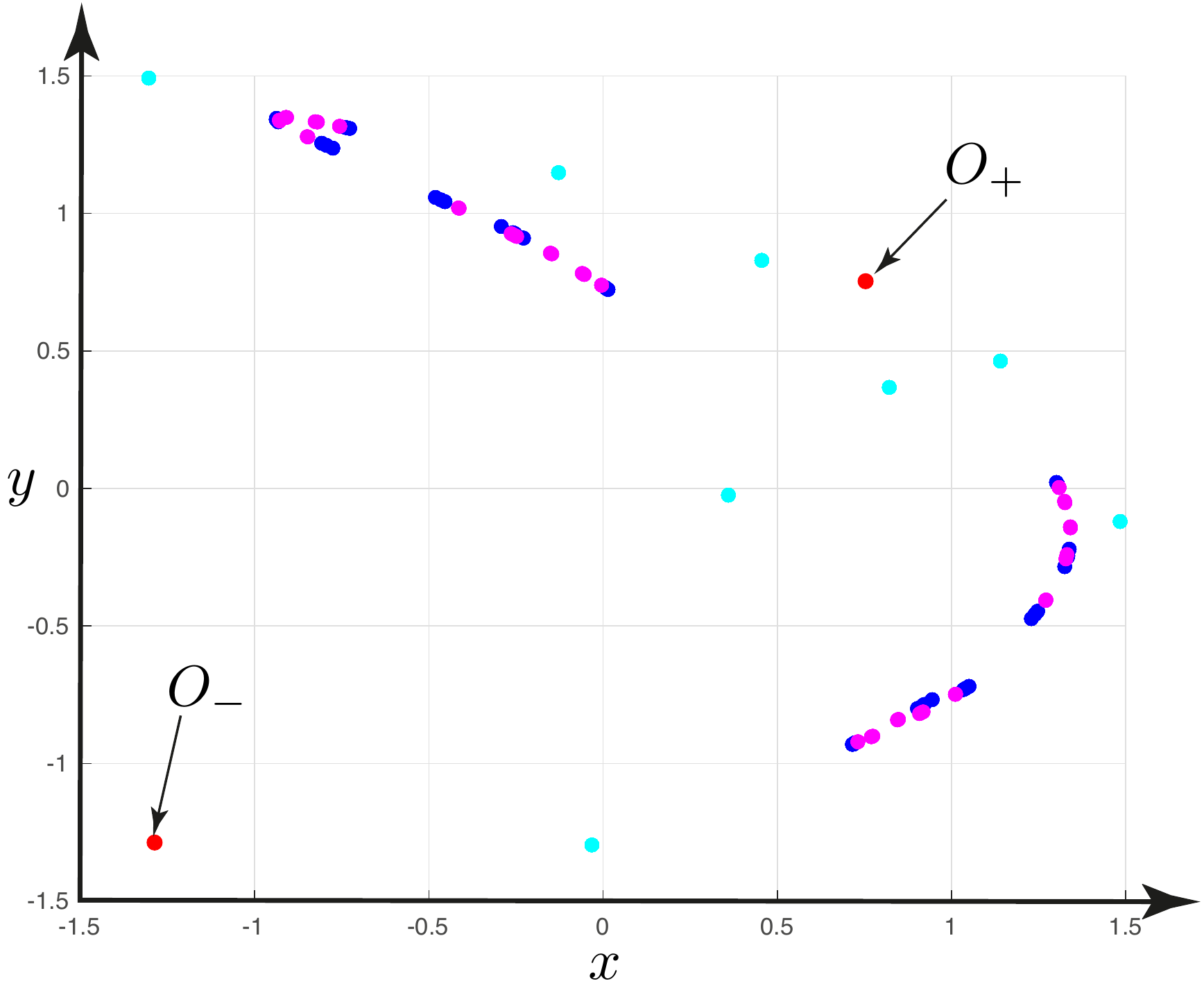}
  }
 \caption{\label{fig:henon:multistab}
 Multistability in the H\'{e}non system \eqref{sys:map}
 with three co-existing self-excited attractors.
 }
\end{figure*}

In \cite{GaliasT-2013} the multistability in the H\'{e}non system was studied for positive parameters. In Fig.~\ref{fig:henon:multistab:3sinks}, for parameters $a = 0.98$, $b = 0.4415$
there are three co-existing self-excited attractors:
period-8 orbit self-excited with respect to $O_\pm$
(used initial data $u_0 = O_+ + \delta \tfrac{\nu^-(O_+)^*}{||\nu^-(O_+)||}$, $\delta = 10^{-4}$),
period-12 orbit self-excited with respect to $O_\pm$
(used initial data $u_0 = O_- + \delta \tfrac{\nu^+(O_-)^*}{||\nu^+(O_-)||}$, $\delta = 0.1$),
and period-20 orbit self-excited with respect to $O_\pm$
(used initial data $u_0 = O_+ - \delta \tfrac{\nu^-(O_+)^*}{||\nu^-(O_+)||}$, $\delta = 0.1$).
In Fig.~\ref{fig:henon:multistab:2sinks-1chaos},
for parameters $a = 0.972$, $b = 0.4575$ there are three co-existing attractors:
period-12 orbit self-excited with respect to $O_\pm$
(used initial data $u_0 = O_- + \delta \tfrac{\nu^+(O_-)^*}{||\nu^+(O_-)||}$, $\delta = 0.01$),
period-16 orbit self-excited with respect to $O_\pm$
(used initial data $u_0 = O_+ - \delta \tfrac{\nu^-(O_+)^*}{||\nu^-(O_+)||}$, $\delta = 10^{-8}$),
and chaotic attractor self-excited with respect to $O_\pm$
(used initial data $u_0 = O_+ + \delta \tfrac{\nu^-(O_+)^*}{||\nu^-(O_+)||}$, $\delta = 0.1$).
Last but no least, in Fig.~\ref{fig:henon:multistab:1sink-2chaos},
for parameters $a = 0.97$, $b = 0.466$ there are three co-existing self-excited attractors:
period-8 orbit self-excited with respect to $O_\pm$
(used initial data $u_0 = O_- + \delta \tfrac{\nu^+(O_-)^*}{||\nu^+(O_-)||}$, $\delta = 0.01$),
and two chaotic attractors each one self-excited with respect to $O_\pm$
(used initial data $u_0 = O_+ + \delta \tfrac{\nu^-(O_+)^*}{||\nu^-(O_+)||}$, $\delta = 10^{-7}$, and
$u_0 = O_+ - \delta \tfrac{\nu^-(O_+)^*}{||\nu^-(O_+)||}$, $\delta = 0.1$, respectively).
In \cite{FalcoliniTL-2013,FalcoliniTL-2016}
the coexistence of periodic orbits
is studied near the critical cases when $b \to \pm 1$ and the map becomes less and less dissipative.
The possible existence of hidden chaotic attractors in the H\'{e}non map
requires further investigation.


In the numerical computation of trajectory over a finite-time interval
it is often difficult to distinguish a \emph{sustained oscillation} from
a \emph{transient oscillation}
(a transient set in the phase space, e.g. chaotic or quasi periodic,
which can nevertheless persist for a long time)
\cite{GrebogiOY-1983,LaiT-2011}.
Thus, a similar to the above classification can be introduced for
the transient sets.

\begin{definition}[\cite{DancaK-2017-CSF,ChenKLM-2017-IJBC}]
 A \emph{transient oscillating set} is called \emph{hidden}
 if it does not involve and attract trajectories
 from a small neighborhood of equilibria;
 otherwise, it is called \emph{self-excited}.
\end{definition}

In the H\'{e}non system~\eqref{sys:map} it is possible to observe
the long-lived transient chaotic sets for $b = 0.3$ and $a \in (1.29, 1.3)$.
For example, for $a = 1.29915$ it is possible to localize a
self-excited transient chaotic set\footnote{
  Similar behavior with hidden transient chaotic set can be obtained in
  various Lorenz-like systems, e.g. in the classical Lorenz system~\cite{ChenKLM-2017-IJBC},
  in the Rabinovich system~\cite{KuznetsovLMPS-2017-arXiv}
} with respect to the saddle $O_+$ ($|\lambda^+(O_+)| < 1$, $|\lambda^-(O_+)| > 1$)
using the initial data
\[
  u_0 = O_+ + \delta\frac{\nu^-(O_+)^*}{||\nu^-(O_+)||}, \quad
  \delta = 0.1,
\]
which persists for $\approx 7300$ iterations and after that contracts into
a period $7$ limit cycle.

In order to distinguish an attracting chaotic set (attractor)
from a transient chaotic set in numerical experiments,
one can consider a grid of points in a small neighborhood of the set
and check the attraction of corresponding trajectories towards the set.
Various examples of hidden transient chaotic sets localization are discussed,
e.g., in \cite{DangLBW-2015-HA,Danca-2016-HA,YuanYW-2017-HA,DancaK-2017-CSF,ChenKLM-2017-IJBC}.

In \cite{Feit-1978}, for parameters $a > 0$ and $b \in (0,1)$,
it is suggested an analytical bounded localization of attractors in the H\'{e}non map
by the set $\mathcal{B} = \overline{M} \setminus (Q \bigcup R_1 \bigcup R_2)$,
where
\begin{equation}\label{eq:setQ}
\begin{aligned}
  & M = \{x,y~|~ x < m, y < m + a\}, \ m = \tfrac{a (1 + 2b)}{1 - b} > 0,
  \\ &
  Q = \{x,y~|~ x < r, y < 0\}, \ r = -\tfrac{1 + \sqrt{1 + 4a}}{2} < 0,
  \\ &
  R_1 = \{x,y~|~ x < -\sqrt{b(m + a) + a - r}, \ y \leq m + a\},
  \\ &
  R_2 = \{x,y~|~ x \leq m, \ y < - \tfrac{1}{b}\big(a + \sqrt{a + bm - r}\big)\}.
\end{aligned}
\end{equation}
A similar set can be considered for negative values of $b$.

Further by $K$ we denote a bounded closed invariant set,
e.g. a maximum attractor with respect to the set of all nondivergent points
from $\mathcal{B}$ (i.e. $u_0 \in \mathcal{B}: \limsup_{t \to \infty}|\varphi^t(u_0)| \neq \infty$).

\section{Finite-time Lyapunov dimension and algorithms for its computation}
\label{sec:FTLE}

The concept of the Lyapunov dimension was suggested
in the seminal paper by Kaplan \& Yorke \cite{KaplanY-1979}
and later it has been developed and rigorously justified in a number of papers.
Nowadays, various approaches to the Lyapunov dimension definition are used
(see, e.g. \cite{Ledrappier-1981,EdenFT-1991}).
Below we consider the concept of the \emph{finite-time Lyapunov dimension}
\cite{Kuznetsov-2016-PLA,KuznetsovLMPS-2017-arXiv},
which is convenient for carrying out numerical experiments with finite time.

Let a nonempty closed bounded set $K \subset \mathbb{R}^2$
be invariant with respect to dynamical system generated by \eqref{sys:map}
$\{\varphi^t\}_{t\geq0}$, i.e. $\varphi^t(K) = K$ for all $t \geq 0$
(e.g. $K$ is an attractor).
Further we use compact notations for
the \emph{finite-time local Lyapunov dimension}:
$\dim_{\rm L}(t,u_0) = \dim_{\rm L}(\varphi^t, u_0)$,
the \emph{finite-time Lyapunov dimension}:
$\dim_{\rm L}(t,K) = \dim_{\rm L}(\varphi^t, K)$,
and for the \emph{Lyapunov dimension}:
$\dim_{\rm L}K = \dim_{\rm L}(\{\varphi^t\}_{t \geq0}, K)$.

Consider linearization of system \eqref{sys:map}
along the solution $u(t,u_0) = \varphi^t(u_0)$, $u_0 \in \mathbb{R}^2$:
\begin{equation}\label{sfl}
  \begin{aligned}
    & v(t+1) = J\big(u(t,u_0)\big) \, v(t), \ v(0) = v_0 \in \mathbb{R}^2,
   \ t \in \mathbb{N}_0.
  \end{aligned}
\end{equation}
Consider a fundamental matrix $\Phi(t,u_0)$ of solutions of linearized system \eqref{sfl}
such that $\Phi(0,u_0)=I$, i.e.
\[
  \Phi(t,u_0) = J(u(t-1,u_0)) J(u(t-2,u_0)) \cdots J(u(1,u_0))J(u_0).
\]
Then for any solution $v(t,v_0)$ of \eqref{sfl} with the initial data  $v(0,v_0)=v_0$
we have
\begin{equation}\label{eq:var}
  v(t,v_0) = \Phi(t,u_0) \, v_0, \quad u_0, v_0 \in \mathbb{R}^2.
\end{equation}
Let $\sigma_i(t,u_0) = \sigma_i(\Phi(t,u_0))$, $i = 1,2$
be the singular values of $\Phi(t,u_0)$
(i.e. $\sigma_i (t, u_0) > 0$ and ${\sigma_i (t, u_0)}^2$ are
the eigenvalues of the symmetric matrix $\Phi(t,u_0)^*\Phi(t,u_0)$
with respect to their algebraic multiplicity),
ordered so that $\sigma_1(t,u_0) \geq \sigma_2(t,u_0) > 0$
for any $t$ and $u_0$.
Consider the ordered set 
of \emph{the finite-time Lyapunov exponents} at the point $u_0$
for $t > 0$:
\begin{equation}\label{ftLE}
\LEs_{1,2}(t,\!u_0)\!=\!\tfrac{1}{t}\ln\sigma_{1,2}(t,\!u_0),\,
\LEs_{1}(t,\!u_0)\!\geq\!\LEs_{2}(t,\!u_0). 
\end{equation}
Consider the \emph{Kaplan-Yorke formula \cite{KaplanY-1979} with respect
to the ordered set} $\lambda_1\geq... \geq \lambda_n$:
\begin{equation}\label{lftKY}
   d^{\rm KY}(\{\lambda_i\}_{i=1}^n)\!=\!
   j+\tfrac{\sum_{i=1}^{j}\lambda_i}{|\lambda{j + 1}|},  \
   j\!=\!\max\{m\!: \sum_{i=1}^{m}\!\lambda_i\!\geq\!0\}.
\end{equation}
For the ordered set of finite-time Lyapunov exponents $\{\LEs_i(t,u_0)\}_{i=1}^2$
and $j(t,u_0)\!=\!\max\{m\!: \sum_{i=1}^{m}\!\LEs_i(t,u_0)\!\geq\!0\}$ we get
\[ 
   d^{\rm KY}\!(\{\!\LEs_i(t,\!u_0)\}_{i=1}^2\!)
   \!=\!\left\{
   \begin{aligned}
     & 0, & j(t,u_0)\!=\!0 \\
     & 1\!+\!\tfrac{\LEs_1(t,u_0)}{|\LEs_2(t,u_0)|}, & j(t,u_0)\!=\!1\\
     & 2 & j(t,u_0)\!=\!2
   \end{aligned}
   \right.
\] 
Then for a certain point $u_0$ and invariant closed bounded set $K$
the \emph{finite-time local Lyapunov dimension}
\cite{Kuznetsov-2016-PLA,KuznetsovLMPS-2017-arXiv} is defined  as
\[
   \dim_{\rm L}(t,u_0) = d^{\rm KY}(\{\LEs_i(t,u_0)\}_{i=1}^2).
\]
and the \emph{finite-time Lyapunov dimension} is as follows
\begin{equation}\label{DOmaptmax}
  \dim_{\rm L}(t, K) = \sup\limits_{u_0 \in K} \dim_{\rm L}(t,u_0).
\end{equation}
In this approach the use of Kaplan-Yorke formula \eqref{lftKY} with
the finite-time Lyapunov exponents can be rigorously justified
by the \emph{Douady--Oesterl\'{e} theorem} \cite{DouadyO-1980},
which implies that for any fixed $t > 0$
the Lyapunov dimension of the map $\varphi^t$ with respect
to a closed bounded invariant set $K$, defined by \eqref{DOmaptmax},
is an upper estimate of the Hausdorff dimension of the set $K$:
\[
  \dim_{\rm H}K \leq \dim_{\rm L}(t, K).
\]

\subsection{Adaptive algorithm for the computation of the finite-time Lyapunov dimension}

To compute the finite-time Lyapunov exponents \eqref{ftLE}
one has to find the fundamental matrix $\Phi(t,u_0)$
  of \eqref{sfl} from the following variational equation
  \begin{equation}\label{vareq}
  \begin{aligned}
      & \!u(s\!+\!1,u_0)\!=\!\varphi(u(s,u_0)), \,\, u(0,u_0)\!=\!u_0,\, s\!=\!0,1,..,t\!-\!1 \\
      & \!\Phi(s\!+\!1,u_0)\!=\!J(u(s,u_0))\Phi(s,u_0),
       \quad \Phi(0,u_0)\!=\!I,
  \end{aligned}
  \end{equation}
  and its \emph{Singular Value Decomposition} (SVD)
  \[
    \Phi(t,u_0) \overset{\text{SVD}}{=}U(t,u_0){\rm \Sigma}(t,u_0){\rm V}^*(t,u_0),
  \]
  where $U(t,u_0)^*U(t,u_0) \equiv I \equiv {\rm V}(t,u_0)^*{\rm V}(t,u_0)$,
  ${\rm \Sigma}(t,u_0)=\text{\rm diag}\{\sigma_1(t,u_0),\sigma_2(t,u_0)\}$
  is a diagonal matrix composed by the \emph{singular values} of $\Phi(t,u_0)$,
  and  compute the finite-time Lyapunov exponents $\LEs_{1,2}(t,u_0)$
  from ${\rm \Sigma}(t,\,u_0)$ as in \eqref{ftLE}.

  To avoid the exponential growth of values in the computation,
  we use the QR factorization and treppeniteration routine:
  \[
  \begin{aligned}
    \!& \!\!\Phi(t,u_0)\!=\!J(u(t-1,u_0)) \cdots J(u(1,u_0)) \boxed{J(u_0)} =
    \\\!&
    \!\!\!=\!J(u(t\!-\!1,u_0))\!\cdots\!\boxed{\!J(u(1,u_0)) Q^0_1\!} R^0_1\!
    =\!\cdots\! \\ &
    \overset{\text{QR}}{=} \overbrace{\!\!Q^0_t\!\!}^{Q} \,\overbrace{\!R^0_t\!\dots\!R^0_1\!}^{R}.
  \end{aligned}
  \]
  Then matrix
  $
   \Sigma(t,u_0)\!=\!U^*(t,u_0) \,\Phi(t,u_0) \, V(t,u_0)
  $
  can be approximated by sequential QR decomposition of the product of matrices:
  \begin{align*}
    \Sigma^0 &:=\!\Phi(t,u_0)^* Q^0_t = (R^0_1)^* \dots (R^0_t)^*
    \overset{\text{QR}}{=} Q^1_t R^1_t \dots R^1_1,& \\
    \Sigma^1 &:=\!(Q^0_t)^* \Phi(t,u_0) Q^1_t =
    (R^1_1)^* \dots (R^1_t)^*
    \overset{\text{QR}}{=} Q^2_t R^2_t \dots R^2_1,& \\
    &\vdots\\
    \Sigma^p &:=\!
      \left\{
      \begin{array}{ll}
        \!(V^p)^* \Phi(t,u_0)^* U^p, & \text{\small($p$ is even)}\\
        \!(U^p)^* \Phi(t,u_0) \ V^p, & \text{\small($p$ is odd)}
      \end{array}
      \right. = &\\
    & \qquad \qquad \qquad \qquad \qquad
    = (R^{p}_1)^*\!\dots\!(R^{p}_t)^*
    \!=\!\left(
    \begin{matrix}
      \sigma_1^{1} & 0 \\
      \cdot & \sigma_2^{j}
    \end{matrix}
    \right),&
  \end{align*}
  where $U^p := Q^0_t Q^2_t \dots Q^{2\lfloor \sfrac{p}{2}\rfloor}_t$,
 $V^p := Q^1_t Q^3_t \dots Q^{2\lceil \sfrac{p}{2} \rceil - 1}_t$,
 and \cite{RutishauserS-1963,Stewart-1997}
 \[
  \sigma_i^{p} = R^{p}_1[i,i] \dots R^{p}_t[i,i] \underset{p \to \infty}{\longrightarrow} \sigma_i(t,u_0), \quad i=1,2.
  \]

  For a large $t$ the convergence can be very rapid~\cite[p.~44]{Stewart-1997}
  (e.g. $p=1$ is taken in \cite[p.~44]{Stewart-1997} for the Lorenz system with the classical parameters).

\begin{figure*}[htp]
 \centering
 \subfloat[Simulation follows~\eqref{LEapprox} with $p = 0$.]{
    \label{fig:henon:2chaos:FTLE_It1}
    \includegraphics[width=0.49\textwidth]{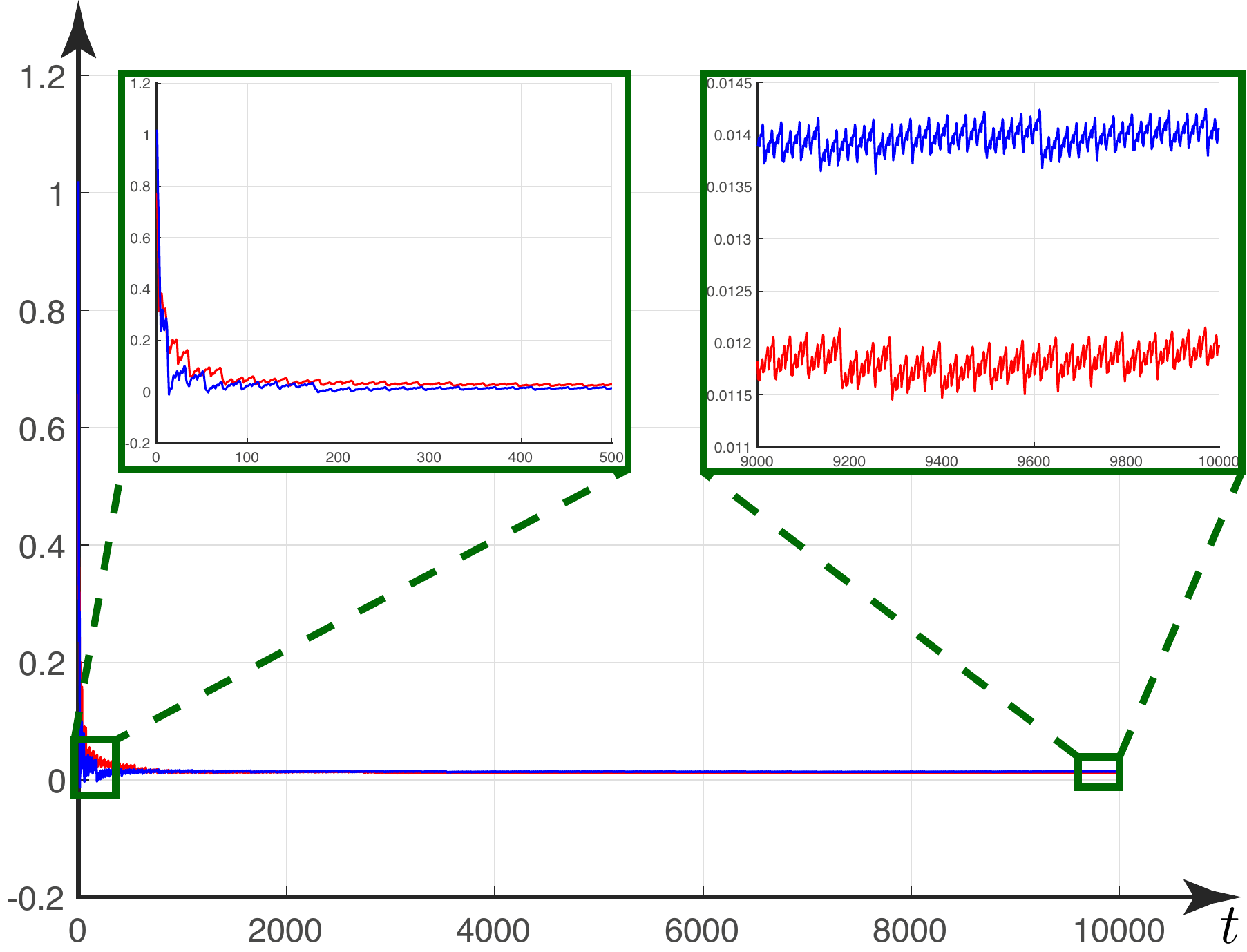}
  }~
  \subfloat[Simulation follows~\eqref{LEapprox} with adaptive value $p(\delta)$, $\delta = 10^{-8}.$]{
    \label{fig:henon:2chaos:FTLE_Tol}
    \includegraphics[width=0.49\textwidth]{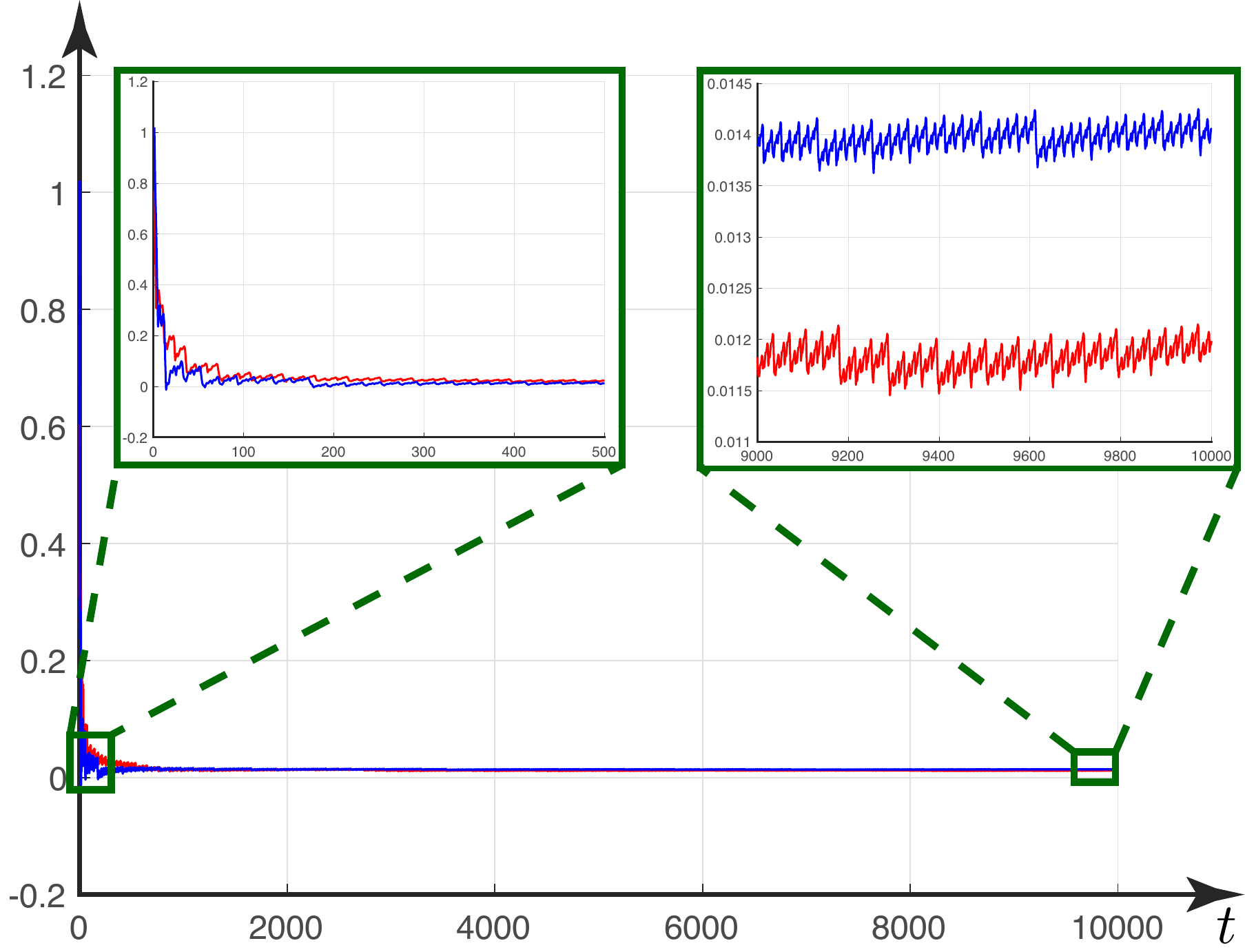}
  }
 \caption{\label{fig:henon:2chaos:FTLE}
 Evolution of $\LEs_1(t,u_0^{\rm a1})$ (red) and $\LEs_1(t,u_0^{\rm a2})$ (blue),
 computed along the trajectories $u(t, u_0^{\rm a1})$ and $u(t, u_0^{\rm a2})$
 for $t \in [1, 10000]$ 
 in the H\'{e}non system \eqref{sys:map}
 with parameters $a = 0.97$,$b = 0.466$. 
 }
\end{figure*}

  For the study of dynamics of the finite-time Lyapunov exponents \cite{KuznetsovLMPS-2017-arXiv}
  we can adaptively choose $p$ for $s\!=\!1,...,t$ so as
  to obtain a uniform estimate with respect to $s$:
  \begin{equation}\label{LEapproxuni}
    \!\!p(\delta)\!=\!p(s,\delta)\!:\,\max_{i=1,2}|
    \tfrac{1}{s}\ln\sigma_i^{p(s)}(s, u_0)-\tfrac{1}{s}\ln\sigma_i^{p(s)\!-\!1}(s,u_0)|\!<\!\delta.
  \end{equation}
  Thus, 
  the finite-time Lyapunov exponents can be approximated\footnote{
     In \emph{Benettin's algorithm}~\cite{BenettinGGS-1980-Part2}
     the so-called
     \emph{finite-time Lyapunov characteristic exponents} (LCEs) \cite{Lyapunov-1892},
     which are the exponential growth rates of norms of the fundamental matrix columns
     $\Phi(t,u_0) = \{v_1(t,u_0),v_2(t,u_0)\}$:
     \(
        \LCEs_{1,2}(t,u_0)= \frac{1}{t}\ln||v_{1,2}(t,u_0)||,
     \)
      are computed by \eqref{LEapprox} with $p=0$:
    \(
      \!\LCEs_{1,2}(t, u_0)\!\approx\! \LEs_{1,2}^{0}(t, u_0) =
      \frac{1}{t}\!\sum_{s=1}^{t} \ln R_{s}^0[i,i].
    \)
    The following artificial analytical example
    demonstrates possible differences between LEs and LCEs:
    the matrix \cite{Kuznetsov-2016-PLA,KuznetsovAL-2016,KuznetsovLMPS-2017-arXiv}
    \(
      R(t)\!=\!\left(\!\!
        \begin{array}{cc}
          1 & e^{at}-e^{-at}\\
          0 & 1 \\
        \end{array}
      \!\!\right)
    \)
    has
    $\LEs_{1,2}(t) = \pm |a|$,
    $
     \LCEs_1(t)\!=\!\tfrac{1}{t} \ln\!\big((e^{at}\!-\tfrac{1}{e^{at}})^2+1\big)^{\tfrac{1}{2}}
     \!\in\!(0, a], \LCEs_2(t) \equiv 0;
    $
    The approximation by Benettin's algorithm becomes worse with increasing time:
    $\LCEs_1(t) \underset{t \to +0}{\longrightarrow} 0 \equiv  \LEs_1^{0}(t), \
      \LCEs_1(t) \!\underset{t \to +\infty}{\longrightarrow}\! a.
    $
    Remark that the notions of LCEs and LEs
    often do not differ (see, e.g. Eckmann \& Ruelle \cite[p.620,p.650]{EckmannR-1985},
    Wolf~et~al.~\cite[p.286,p.290-291]{WolfSSV-1985},
    and Abarbanel~et~al.~\cite[p.1363,p.1364]{AbarbanelBST-1993}),
    e.g. relying on ergodicity,
    however, the computations of LCEs by \eqref{LEapprox}
    and LEs by $\LEs^{0}(t, u_0)$ may give \emph{non~relevant results}.
 }
  as
  \begin{equation}\label{LEapprox}
  \!\LEs_i(t, u_0)\!\approx\!
    \LEs_i^{p(\delta)}(t, u_0)\!= 
    \tfrac{1}{t}\!\sum_{s=1}^{t} \ln R_{s}^{p(\delta)}[i,i].
  \end{equation}

\begin{figure}[t]
    \centering
    \includegraphics[width=0.49\textwidth]{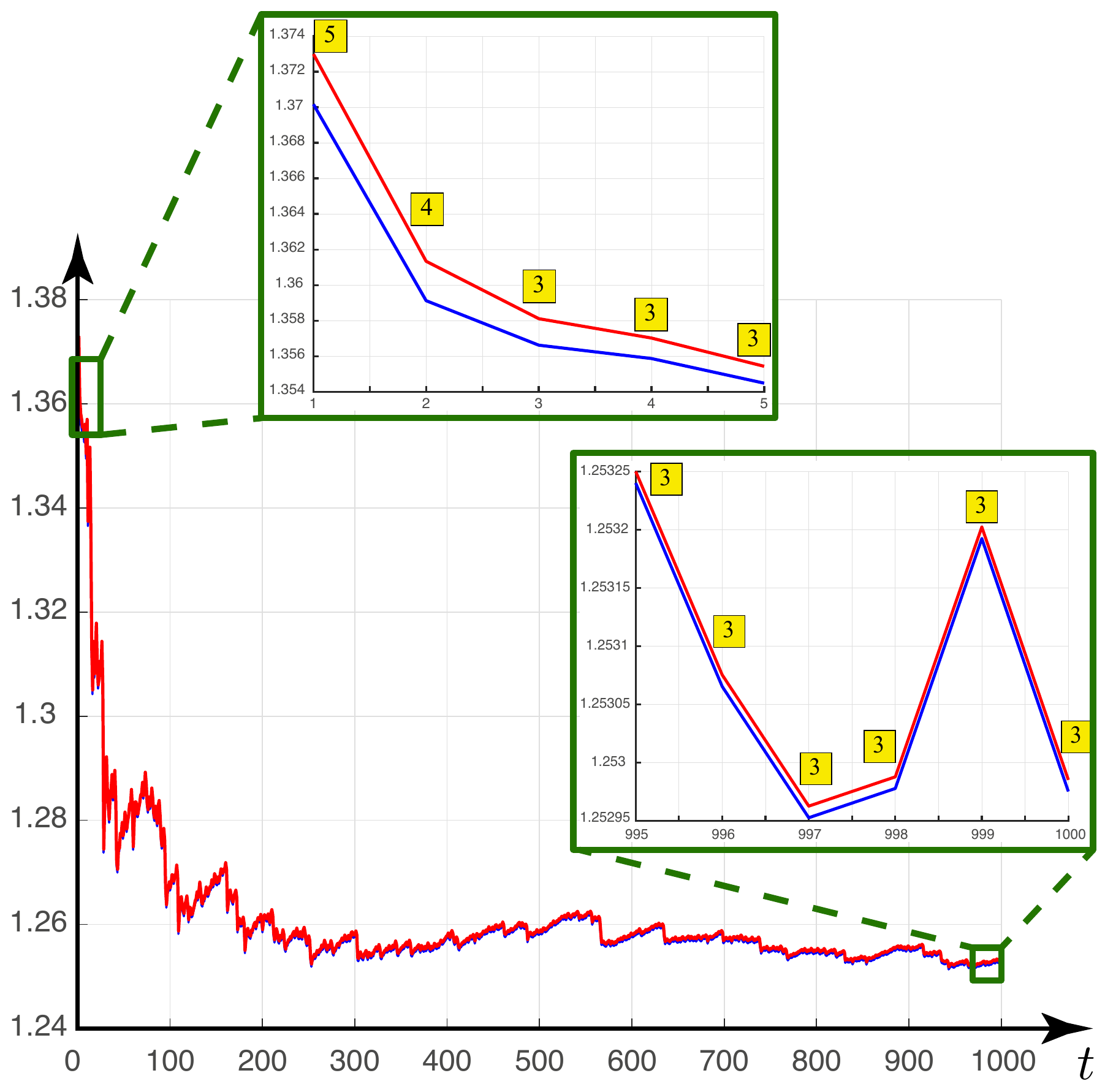}
    \caption{\label{fig:henon:FTLD}
    Evolution of the finite-time local Lyapunov dimension $\dim_{\rm L}(t,u_0)$
    computed with $p=0$ (blue) and by the adaptive algorithm (red).
    The corresponding values in yellow box shows
    the number of iterations $p(t,\delta)$, which are
    necessary to meet the tolerance $\delta = 10^{-8}$.
    }
  \end{figure}

 For the H\'{e}non system \eqref{sys:map} with canonical parameters $a = 1.4$, $b = 0.3$,
 using the described above adaptive algorithm with $\delta = 10^{-8}$,
 we calculate the finite-time local Lyapunov dimension $\dim_{\rm L}(t,u_0)$,
 where $u_0$ is the initial point of the trajectory $u(t,u_0)$ that localizes
 the self-excited attractor (see Fig.~\ref{fig:henon:attr:SE}).
 The comparison of the graphics for the adaptive algorithm and
 the algorithm with $p=0$
 is presented in Fig.~\ref{fig:henon:FTLD}.
 For
 $t = 1000$
 we obtain
 $
 \dim_{\rm L}(1000, u_0) \approx 1.253$.
 The corresponding numerical routine implemented in MATLAB is presented
 in Appendix~\ref{appendix:matlab_code}.

Applying the statistical physics approach and assuming the ergodicity
(see, e.g. \cite{KaplanY-1979,Ledrappier-1981,FredericksonKYY-1983}),
the Lyapunov dimension of attractor
$\dim_{\rm L} K$ 
is often estimated by the local Lyapunov dimension
$\dim_{\rm L} (t, u_0)$,
corresponding to a
``typical'' trajectory, which belongs to the attractor:
$\{u(t,u_0), t \geq 0 \},\ u_0 \in K$,
and its limit value $\lim_{t\to+\infty}\dim_{\rm L} (t, u_0)$.
However, rigorous check of ergodicity for the Henon system
with a particular value of the parameters
is a challenging task (see, e.g. \cite{BenedicksC-1991,BenedicksY-1993}).
See, also related discussions in \cite{BarreiraS-2000}\cite[p.118]{ChaosBook}\cite{OttY-2008}\cite[p.9]{Young-2013}
\cite[p.19]{PikovskyP-2016}, and the works \cite{KuznetsovL-2005,LeonovK-2007}
on the \emph{Perron effects of the largest Lyapunov exponent sign reversals}.
For example, consider parameters $a = 0.97$,$b = 0.466$ (see Fig.~\ref{fig:henon:multistab:1sink-2chaos}).
In this case for $u_0^1=O_- + \delta_1 \tfrac{\nu^+(O_-)^*}{||\nu^+(O_-)||}$, $\delta_1 = 10^{-4}$
and $u_0^2 = O_+ - \delta_2 \tfrac{\nu^-(O_+)^*}{||\nu^-(O_+)||}$, $\delta_2 = 0.1$
after a transient process during $[0,T_{\rm trans}=10^5]$
we get initial points $u_0^{\rm a1}$ and $u_0^{\rm a2}$, respectively
and compute finite-time Lyapunov exponents and finite-time local Lyapunov dimension
for the time interval $[0,T= 10^{4}]$
by the adaptive algorithm with $\delta = 10^{-8}$:
dynamics of the finite-time Lyapunov exponents and finite-time local Lyapunov dimension
is presented in  Fig.~\ref{fig:henon:2chaos:FTLE}, finally we have
\[
  \begin{aligned}
  &\LEs_1(T, u_0^{\rm a1})\!\approx\!0.01198317,
  \ \dim_{\rm L}(T, u_0^{\rm a1})\!\approx\!1.01545113, \\
  &\LEs_1(T, u_0^{\rm a2})\!\approx\!0.01405416,
  \ \dim_{\rm L}(T, u_0^{\rm a2})\!\approx\!1.01807321.
  \end{aligned}
\]
%

\begin{figure}[h]
    \centering
    \includegraphics[width=0.5\textwidth]{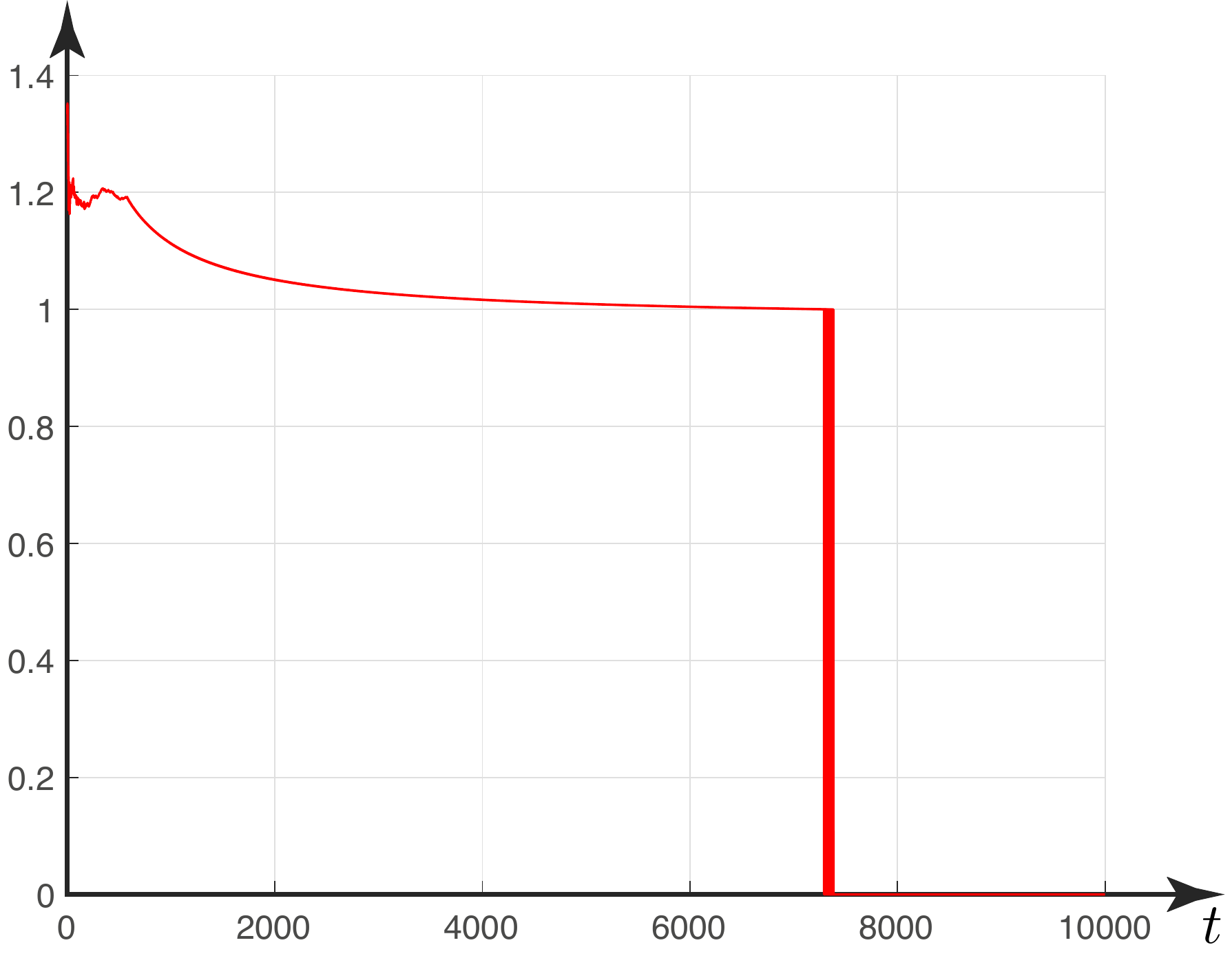}
    \caption{\label{fig:henon:transFTLD}
    Evolution of the finite-time local Lyapunov dimension $\dim_{\rm L}(t,u_0)$ (red)
    computed via the adaptive algorithm along the transient chaotic set for
    parameters $a = 1.29915$, $b = 0.3$ and $t \in [1,10000]$,
    and tolerance $\delta = 10^{-8}$.
    For $t \in [1,7300]$ the behavior seems to be chaotic.
    }
  \end{figure}

In one of the pioneering works by Yorke~et~al.~\cite[p.190]{FredericksonKYY-1983}
the \emph{exact} limit values of finite-time Lyapunov exponents,
if they exist and are the same for all $u \in K$,
are called the \emph{absolute} ones,
and it is noted that the \emph{absolute Lyapunov exponents} \emph{rarely exist}.
Remark that while the time series obtained from a \emph{physical experiment}
are assumed to be reliable on the whole considered time interval,
the time series, obtained numerically from \emph{mathematical dynamical model},
can be reliable on a limited time interval only due to computational errors.
Also, if the trajectory belongs to a transient chaotic set (see, e.g. Fig.~\ref{fig:henon:transient}), which can be (almost) indistinguishable numerically from sustained chaos, then any very long-time computation may be insufficient
to reveal the limit values of the finite-time Lyapunov exponents
and finite-time Lyapunov dimension (see Fig.~\ref{fig:henon:transFTLD})\cite{KuznetsovLMPS-2017-arXiv}.

Thus, to get a reliable numerical estimation of the Lyapunov dimension of attractor $K$
we localize the attractor $K \subset K^{\varepsilon}$,
consider a grid of points $K^{\varepsilon}_{\rm grid}$ on $K^{\varepsilon}$,
and find the maximum of the corresponding finite-time local Lyapunov dimensions
for a certain time interval $[0,T]$:
\begin{equation}\label{dimLmunest}
\begin{aligned}
  \dim_{\rm H}K
  \leq
  \dim_{\rm L}K
  \approx \inf_{t \in [0,T]} \max_{u \in K^{\varepsilon}_{\rm grid}} \dim_{\rm L}(t, u) \\
 =\inf_{t \in [0,T]}\max\limits_{u \in K^{\varepsilon}_{\rm grid}}
 \left(j(t,u) + \frac{\LEs_{j(t,u)}(t,u)}{|\LEs_{j(t,u)+1}(t,u)|}
 \right)\\
  \leq
  \max_{u \in K^{\varepsilon}_{\rm grid}} \dim_{\rm L}(T, u) \approx \dim_{\rm L}(T,K).
\end{aligned}
\end{equation}
Additionally, we can consider a set $K^{\varepsilon}_{\rm rnd}$ of
$N(\varepsilon)$ random points in $K^{\varepsilon}$,
where $N(\varepsilon)$ is the number points in $K^{\varepsilon}_{\rm grid}$.
If the maximum of the finite-time local Lyapunov dimensions for
$K^{\varepsilon}_{\rm rnd}$ and $K^{\varepsilon}_{\rm grid}$ are different,
i.e.
 $|\max\limits_{u \in K^{\varepsilon}_{\rm rnd}} \dim_{\rm L}(\varphi^T,u)-
\max\limits_{u \in K^{\varepsilon}_{\rm grid}} \dim_{\rm L}(\varphi^T,u)|>\delta$,
then we decrease $\varepsilon$.
This may help to improve reliability of the result
and at the same time to ensure its repeatability.

\section{The Lyapunov dimension: analytical estimations and exact value}
\label{sec:LD_exact}

To estimate the Hausdorff dimension of invariant closed bounded set $K$,
one can use the map $\varphi^t$ with any time $t$
(e.g., $t = 0$ leads to the trivial estimate $\dim_{\rm H}K \leq 2$),
and, thus, the best estimation is
\[
  \dim_{\rm H}{K} \le \inf_{t\geq0}\dim_{\rm L} (t, K).
\]
The following property:
\begin{equation}\label{DOlim}
  \inf_{t\geq0}\sup\limits_{u_0 \in K} \dim_{\rm L}(t,u_0)
  = \liminf_{t \to +\infty}\sup\limits_{u_0 \in K} \dim_{\rm L}(t,u_0),
\end{equation}
allows one to introduce the \emph{Lyapunov dimension}  of $K$
as \cite{Kuznetsov-2016-PLA}
\begin{equation}\label{DOinf}
  \dim_{\rm L} K 
  = \liminf_{t \to +\infty}\sup\limits_{u_0 \in K} \dim_{\rm L}(t,u_0)
\end{equation}
and get an upper estimation of the Hausdorff dimension:
\[
  \dim_{\rm H}{K} \le \dim_{\rm L} K.
\]
Recall that a set with noninterger Hausdorff dimension
is referred as a \emph{fractal set} \cite{EckmannR-1985}.

In contrast to the finite-time Lyapunov dimension \eqref{DOmaptmax},
the Lyapunov dimension \eqref{DOinf}\footnote{
This definition can be reformulated via the singular value function
$\omega_d(D\varphi^t(u)) =
  \sigma_1(t, u)\cdots\sigma_{\lfloor d \rfloor}(t, u)
    \sigma_{\lfloor d \rfloor+1}(t, u)^{d-\lfloor d \rfloor},
$ where $d \in [0,n]$ and ${\lfloor d \rfloor}$ is the largest integer less or equal to $d$: \\
\(
  \dim_{\rm L}(t,u)\!=\!\!\max\{d\!\in\![0,n]\!: \omega_{d}(D\varphi^t(u))\!\geq\!1 \}
\)
and $\dim_{\rm L}K  = \liminf\limits_{t \to +\infty}\sup\limits_{u \in K}\dim_{\rm L}(t,u)$.
  Another approach to the introduction of the Lyapunov dimension of dynamical system
  was developed by Constantin, Eden, Foia\c{s}, and Temam
  \cite{ConstantinFT-1985,Eden-1990,EdenFT-1991}.
  They consider $\big(\omega_{d}(D\varphi^t(u))\big)^{1/t}$
  instead of $\omega_{d}(D\varphi^t(u))$
  and apply the theory of positive operators
  to prove the existence of a critical point
  $u^{cr}_{\rm E}$ (which may be not unique),
  where the corresponding global Lyapunov dimension
  achieves maximum (see \cite{Eden-1990}):
  \(
  \dim_{\rm L}^{\rm E}(\{\varphi^t\}_{t\geq0},K)
  =\inf\{d \in [0,n]:
  \lim\limits_{t \to +\infty} \max\limits_{u \in K}\ln\big(\omega_{d}(D\varphi^t(u))\big)^{1/t}<0\} =
  \inf\{d \in [0,n]:
  \limsup\limits_{t \to +\infty}\ln\big(\omega_{d}(D\varphi^t(u^{cr}_{\rm E}))\big)^{1/t}<0\}
  =
  \dim_{\rm L}^{\rm E}(\{\varphi^t\}_{t\geq0},u^{cr}_{\rm E}),
  \)
  and, thus, rigorously justify the usage of the local Lyapunov dimension
  $\dim_{\rm L}^{\rm E}(\{\varphi^t\}_{t\geq0},u)$.
  However this definition does not have a clear sense
  for finite time.
}
is \emph{invariant under smooth change of coordinates} \cite{KuznetsovAL-2016,Kuznetsov-2016-PLA}.
This property and a proper choice of smooth change of coordinates
may significantly simplify the computation of the Lyapunov dimension of dynamical system.
Consider an effective analytical approach, proposed by Leonov \cite{Leonov-1991-Vest,LeonovB-1992,Kuznetsov-2016-PLA},
for estimating the Lyapunov dimension. In the work \cite{Kuznetsov-2016-PLA} it is shown how this approach can be justified
by the invariance of the Lyapunov dimension of compact invariant set
with respect to the special smooth change of variables $w=h(u)$
with $Dh(u)=e^{V(u)(j+s)^{-1}}S$, where $V(u)$
is a continuous scalar function
and $S$ is a nonsingular matrix.
Let $\sigma_i(J(u_0))$, $i=1,2$
be the singular values of $J(u_0)$
(i.e. the square roots of the eigenvalues of the symmetrized Jacobian matrix
$J(u_0)^*J(u_0)$),
ordered so that $\sigma_1(J(u_0)) \geq \sigma_2(J(u_0)) > 0$
for any $u_0 \in K$.

\begin{theorem}[\cite{Kuznetsov-2016-PLA}]\label{thm:LD-estimate-Vdt}
If there exist 
a real $s \in [0,1]$,
a continuous scalar function $V(u)$,
and a nonsingular $2\times 2$ matrix $S$
such that
\begin{equation}\label{ineq:weilSV}
  \sup_{u_0 \in K}\!\!\big(
  \ln\sigma_1(J(u_0)) + s\ln\sigma_2(J(u_0))
  + \big(V(\varphi(u_0))-V(u_0)\big)\big)\!<\!0,
\end{equation}
then
\[
   \dim_{\rm H}K \leq \dim_{\rm L}K < 1+s.
\]
\end{theorem}
\noindent To avoid numerical localization of attractor,
we can consider estimation \eqref{ineq:weilSV}
e.g. by the absorbing set or the whole phase space.

In \cite{BoichenkoL-1998,PogromskyM-2011} it is demonstrated how a technique similar to the above can be effectively used to derive constructive upper bounds of the topological entropy of dynamical systems.

If we consider $V=0$ and $S=I$,
then we get \emph{the Kaplan-Yorke formula with respect
to the ordered set of logarithms of the singular values of the Jacobian matrix}:
\(
  d_{\rm L}^{\rm KY}(\{\ln\sigma_i(u,S)\}_{i=1}^2),
\)
and its supremum on the set $K$ gives an upper estimation of the finite-time Lyapunov
dimension.
This is a generalization of ideas, discussed  e.g. in \cite{DouadyO-1980,Smith-1986},
on the Hausdorff dimension estimation by the eigenvalues of symmetrized Jacobian matrix.

The Jacobian \eqref{eq:jac} of the H\'{e}non map \eqref{sys:henon2}
has the following singular values:
\begin{align*}
  \sigma_{1}(x) &= \tfrac{1}{2}\left(\sqrt{4 x^2 + (1 + |b|)^2} + \sqrt{4 x^2 + (1 - |b|)^2} \right) \geq 1, \\
  \sigma_{2}(x) &=\frac{|b|}{\sigma_{1}(x)} \leq |b|.
\end{align*}
These expressions give the following estimation:
\[
\dim_{\rm H}K \leq \dim_{\rm L}K \leq
\sup_{(x,y) \in \mathcal{B}}\!\left(1 + \frac{1}{1 - \tfrac{\ln|b|}{\ln \sigma_{1}(x)}}\right).
\]
The maximum of the right-hand side value is determined by the maximum value of $x^2$
(or $|x|$) on $\mathcal{B}$.
In~\cite{Hunt-1996}, for canonical parameters $a = 1.4$, $b = 0.3$
it was considered the square
$(x,y) \in [-1.8, 1.8] \times [-1.8, 1.8]$ that gives
the estimation
\[
  \dim_{\rm H}K \leq \dim_{\rm L}K
  \leq 1 + \frac{1}{1 - \tfrac{\ln0.3}{\ln \sigma_{1}(1.8)}} \approx 1.523
\]
Using Feit's analytical localization
\eqref{eq:setQ} we can get
\begin{multline*}
\dim_{\rm H}K \leq \dim_{\rm L}K \leq \\
\leq 1 +
\left(1 - \tfrac{\ln b}{\ln \sigma_{1}\left(\max\big\{-r,m,\sqrt{b(m + a) + a - r}\big\}\right)}\right)^{-1}.
\end{multline*}
For parameters $a = 1.4$, $b = 0.3$ we obtain
$\dim_{\rm H}K \leq \dim_{\rm L}K \leq 1.5319$.

Remark, that if the Jacobian matrix $J(u_{eq})$
at one of the equilibria has simple real eigenvalues:
$|\lambda_{1}(u_{eq})| \geq |\lambda_{2}(u_{eq})|$,
then the invariance of the Lyapunov dimension
with respect to linear change of variables implies \cite{Kuznetsov-2016-PLA}
the following
\begin{equation}\label{dimLeq}
  \dim_{\rm L}u_{eq}
  =
  d_{\rm L}^{\rm KY}(\{ \ln|\lambda_i(u_{eq})|\}_{i=1}^2).
\end{equation}
If the maximum of local Lyapunov dimensions on the B-attractor,
involving all equilibria, is achieved at equilibrium point:
$\dim_{\rm L} u^{cr}_{eq} = \max_{u_0 \in K} \dim_{\rm L} u_0$,
then this allows one to get analytical formula of
the \emph{exact Lyapunov dimension}\footnote{
This term was suggested by Doering~et~al.~in~\cite{DoeringGHN-1987}.}.
In general, a \emph{conjecture on the Lyapunov dimension of self-excited attractor} \cite{Kuznetsov-2016-PLA,KuznetsovL-2016-ArXiv}
is that the Lyapunov dimension of typical self-excited attractor
does not exceed the Lyapunov dimension of one of unstable equilibria,
the unstable manifold of which intersects with the basin of attraction
and visualize the attractor.

Following \cite{Leonov-2002} for
the H\'{e}non system \eqref{sys:map} with
parameters $a > -\tfrac{(b-1)^2}{4}$ and $|b| < 1$
we can consider
\[
S =
\left(
  \begin{array}{cc}
    1 & 0 \\
    0 & \sqrt{|b|} \\
  \end{array}
\right),
\gamma= \tfrac{1}{(b-1-2x_{-})\sqrt{x_{-}^2+|b|}}, \ s \in [0,1).
\]
In this case we have
\[
SJ\big((x,y)\big)S^{-1} =
\left(
  \begin{array}{cc}
    -2x & \sqrt{|b|} \\
    \sqrt{|b|} & 0 \\
  \end{array}
\right),
\]
and
\[
\sigma_1((x,y),S)\!=\!\big(\sqrt{x^2+|b|}+|x|\big), \
\sigma_2((x,y),S)\!=\!\tfrac{|b|}{\sigma_1((x,y),S)}.
\]
If we take $V((x,y)) = \gamma(1-s)(x+by)$,
then condition \eqref{ineq:weilSV} with $j=1$
and
\[
s > s^* =\dfrac{1}{1 - \tfrac{\ln |b|}{\ln\sigma_1((x_{-},x_{-}),S)}}
\]
is satisfied for all $(x,y) \in \mathbb{R}^2$
and we do not need any localization of the set $K$ in the phase space.
By~\eqref{dimLeq} and \eqref{lftKY}, at the equilibrium point $u^{cr}_{eq} = (x_{-},x_{-})$
we get
\begin{multline*}
  \dim_{\rm L}(\{\varphi^t\}_{t\geq0},(x_{-},x_{-})) = \\ =
  \dim_{\rm L}^{\rm KY}(\{\ln\lambda_i(x_{-},x_{-})\}_1^2)
  =
  1+s^*.
\end{multline*}
Therefore, for a bounded invariant set $K \ni (x_-,x_-)$
(e.g. maximum B-attractor)
we have \cite{Leonov-2002}
\begin{multline*}
\dim_{\rm L}(\{\varphi^t\}_{t\geq0},K^{B}) =
  \dim_{\rm L}(\{\varphi^t\}_{t\geq0},(x_{-},x_{-})) = \\
  =1+\dfrac{1}{1 - \tfrac{\ln |b|}{\ln\sigma_1((x_{-},x_{-}),S)}}.
\end{multline*}
Here for $a=1.4$ and $b=0.3$ we have
$\dim_{\rm L}(\{\varphi^t\}_{t\geq0},K^{B})=1.495\,...$.

Embedding of the attractor into three-dimensional phase space
(see attractors of generalized H\'{e}non map in \cite{BaierK-1990,GonchenkoGS-2010})
increases the Lyapunov dimension by one.

Using the above approach one can obtain the Lyapunov dimension formulas
for invariant sets of other discrete systems (see, e.g. \cite{ReitmannS-2000,LeonovP-2005}).

\section{\label{sec:conclusion} Conclusion}
In this work the H\'{e}non map with
positive and negative values of the shrinking parameter
is considered and transient oscillations, multistability and possible existence of hidden attractors are studied.
A new adaptive algorithm of the finite-time Lyapunov dimension computation
is used for studying the dynamics of the dimension.
Analytical estimate of the Lyapunov dimension using localization of attractors
is given.
The proof of the conjecture on the Lyapunov dimension of self-excited attractors
and derivation of the {exact Lyapunov dimension} formula
are extended to negative values of the parameters.

\section*{Acknowledgements}
The work was supported by Russian Science Foundation project (14-21-00041).


\clearpage
\section{\label{appendix:matlab_code} Appendix: MATLAB code}

\begin{lstlisting}[caption={{\bf henonMap.m} -- function defining the H\'{e}non map.},
    label=lst:henon_map]
function out = henonMap( x, a, b )
  out = zeros(6,1);

  out(1) = a + b * x(2) - x(1)^2;
  out(2) = x(1);

  out(3:6) = [-2 * x(1), b; 1, 0];
end
\end{lstlisting}

\begin{lstlisting}[caption={{\bf qr\_pos.m} -- function implementing
the QR decomposition with positive diagonal elements in R.},
    label=lst:qr_pos]
function [Q, R] = qr_pos(A)

    [Q, R] = qr(A);

    D = diag(sign(diag(R)));
    Q = Q * D; R = D * R;
end
\end{lstlisting}

\begin{lstlisting}[caption={{\bf treppeniterationQR.m} -- function implementing
the treppeniteration QR decomposition for product of matrices.},
    label=lst:treppeniteration_qr]
function [Q, R] = treppeniterationQR(matFact)

    [~, dimOde, nFactors] = size(matFact);

    R = zeros(dimOde, dimOde, nFactors);
    Q = eye(dimOde, dimOde);

    for jFactor = nFactors : -1 : 1
        C = matFact(:, :, jFactor) * Q;
        [Q, R(:, :, jFactor)] = qr_pos(C);
    end
end
\end{lstlisting}

\begin{lstlisting}[caption={{\bf computeLEsDiscrTol.m} -- function implementing
the LEs numerical computation via the approximation of the singular values matrix
with adaptively chosen number of iterations.},
    label=lst:treppeniteration_qr]
function [t, LEs, svdIterations] = computeLEsDiscrTol(extMap, initPoint, nFactors, LEsTol)

% Dimension of the map :
dimMap = length(initPoint);

% Dimension of the ext. map (map+var. eq.):
dimExtMap = dimMap * (dimMap + 1);
initCond = initPoint(:);
fundMat = zeros(dimMap, dimMap, nFactors);

% Main loop : factorization of the fundamental matrix
for iFactor = 1 : nFactors

  extMapSolution = extMap(initCond);

  fundMat(:, :, nFactors-iFactor+1) = reshape(extMapSolution((dimMap + 1) : dimExtMap), dimMap, dimMap);

  initCond = extMapSolution(1 : dimMap);

end

t = 1 : 1 : nFactors;
LEs = zeros(nFactors, dimMap);
svdIterations = zeros(nFactors, 1);

for iFactor = 1 : nFactors

  currFactorization = fundMat(:, :, nFactors-iFactor+1 : nFactors);

  currSvdIteration = 1;

  LEsWithinTol = false;

  while ~LEsWithinTol

    % Save current iteration number:
    svdIterations(iFactor) = currSvdIteration;

    % Calculate current LEs approximation:
    [~, R] = treppeniterationQR(currFactorization);
    for jFactor = 1 : iFactor
      currFactorization(:, :, jFactor) = R(:, :, iFactor-jFactor+1)';
    end
    accumLEs = zeros(1, dimMap);
    for jFactor = 1 : iFactor
      accumLEs = accumLEs + log(diag(currFactorization(:, :, jFactor))');
    end
    LEs(iFactor, :) = accumLEs / iFactor;

    % Compare with previous approximation:
    if currSvdIteration > 1
      LEsWithinTol = all(abs(LEs(iFactor, :) - prevLEs) < LEsTol);
    end

    % Update
    currSvdIteration = currSvdIteration+1;
    prevLEs = LEs(iFactor, :);

  end
end
end
\end{lstlisting}

\begin{lstlisting}[caption={{\bf kaplanYorkeFormula.m} -- function implementing
the local Lyapunov dimension calculation via Kaplan-Yorke formula.},
    label=lst:kaplan_yorke]
function LD = kaplanYorkeFormula(LEs)

% Initialization of the local Lyupunov dimention :
LD = 0;

% Number of LCEs :
nLEs = length(LEs);

% Sorted LCEs :
sortedLEs = sort(LEs, 'descend');

% Main loop :
leSum = sortedLEs(1);
if ( sortedLEs(1) > 0 )
  for i = 1 : nLEs-1
    if sortedLEs(i+1) ~= 0
     LD = i + leSum / abs( sortedLEs(i+1) );
     leSum = leSum + sortedLEs(i+1);
     if leSum < 0
        break;
     end
    end
  end
end
end
\end{lstlisting}

\begin{lstlisting}[caption={{\bf henonLD.m} -- script with application
of the described numerical procedure for LEs computation to the H\'{e}non system.},
    label=lst:henon_LD]
function henonLD

    % Canonical parameters
    a = 1.4; b = 0.3;

    function out = J( x, a, b )
        out = [-2 * x(1), b; 1, 0];
    end

    % Equilibrium
    S1 = 1/2*((b-1) + sqrt((b-1)^2 + 4*a));

    [V1, D1] = eig(J([S1, S1], a, b));
    D1 = diag(D1);

    IX1 = find(abs(D1) > 1);

    % Self-excited attractor with respect to S1
    delta = 1e-3;
    initPoint = [S1, S1] + delta* V1(:, IX1(1))' / norm(V1(:, IX1(1)));

    % Parameters for numerical procedure
    nFactors = 1000;
    LEsTol = 1e-8;

    % LEs computation
    [t, LEs, svdIterations] = computeLEsDiscrTol(@(x) henonMap(x, a, b), initPoint, nFactors, LEsTol);

    % LD computation
    LD = cellfun(@kaplanYorkeFormula, num2cell(LEs, 2));

    % Plotting
    figure(1); hold on;
    plot(t, LEs(:, 1), 'Color', 'red');
    plot(t, LEs(:, 2), 'Color', 'blue');
    hold off;
    grid on; axis on;
    xlabel('t'); ylabel('LE')

    figure(2); hold on;
    plot(t, LD, 'Color', 'green');
    hold off;
    grid on; axis on;
    xlabel('t'); ylabel('LD')
end
\end{lstlisting}

\end{document}